\documentstyle[12pt,epsfig]{article}

\title{\Large{\bf{$\Theta^{+}$ hypernuclei in relativistic mean field model
}}}

\author{  X.H. Zhong,
\,\,Y.H. Tan \, \,L. Li, \,\, P.Z. Ning \\
{Department of Physics, Nankai University,
 Tianjin 300071, P. R. China}}
\begin{document}
\maketitle

\begin{abstract}
We have investigated the properties of $\Theta^{+}$ in nuclei
within the framework of relativistic mean field. The coupling
constants are educed with quark meson coupling model. There is
strong attractive interaction for $\Theta^{+}$-nucleus and
$\Theta^{+}$ can be bind in nuclei. The depth of optical potential
for $\Theta^{+}$ in nuclear matter is estimated.

 PACS: 21.10.Dr, 26.90.Xh,13.75.Ev.
 \\Key words: relativistic mean field, $\Theta^{+}$ hypernuclei

\end{abstract}
\section{Introduction}
Since the extremely narrow positive strangeness pentaquark state
$\Theta^{+}$($uudd\bar{s}$) of mass 1.54 GeV and $J^{P} = 1/2^{+}$
is predicted\cite{a} and with several experimental searches being
successfully undertaken\cite{a1,a2,a3,a4,a5,a6,a7}, pentaquark
becomes the hot topic of investigation. The discovery of the
exotic baryon $\Theta^{+}$ with positive strangeness opens new
possibilities of forming exotic $\Theta^{+}$ hypernuclei which,
like in the case of negative strangeness $\Lambda$, $\Sigma$,
$\Xi$ hypernuclei which have obtained steady progress at the
experimental and theoretical levels\cite{b1,b2,b3,b4,b5}, can
provide information unreachable or complementary to that obtained
in elementary reactions.

Suggestions that $\Theta^{+}$ could be bound in nuclei have
already been made \cite{c1,c2}and the selfenergy of the
$\Theta^{+}$ pentaquark in nuclei is calculated by D. Cabrera $et
al.$\cite{c2}, they show that the in-medium renormalization of the
pion in the two meson cloud of the $\Theta^{+}$ leads to a sizable
attraction, enough to produce a large number of bound and narrow
$\Theta^{+}$ states in nuclei. However, the spin, isospin and
parity of $\Theta^{+}$ are still not very determinate, there are
many discussions about these aspects\cite{a,c1,d1,d2,d3,d5}. In
this work, we assume that $J^{P} = 1/2^{+}$ and $I=0$ for
$\Theta^{+}$ as some suggestions in Refs.\cite{a,c1}, thus we can
study $\Theta^{+}$ hypernuclei in the framework of relativistic
mean field (RMF),which has been  used to investigate $\Lambda$,
$\Sigma$, $\Xi$ hypernuclei\cite{aa2,e1,e2,e3}.

In the present work, our main aim is to determine the coupling
constants $\sigma-\Theta^{+}$, $\omega-\Theta^{+}$ for the RMF
calculation. Then we attempt to investigate the $\Theta^{+}$
hypernuclei in the framework of relativistic mean field, from
which we want to know whether the $\Theta^{+}$ baryon could be
bound in nuclei and the how much the depth of the
$\Theta^{+}$-nucleus potential.

This paper is organized as follows. The framework of hypernuclei
in RMF is introduced in Sec.2. The coupling constants are educed
 and determined in Sec.3. The calculations and analysis is given
 in Sec.4. The last Sec. is the summary.

\section{Hypernuclei in RMF model}
In the RMF theory, the effective Lagrangian density which includes
the nucleon and $\Theta^{+}$-hyperon can be written as
 \begin{eqnarray}
\mathcal{L}=\mathcal{L}_{\mathrm{N}}+\mathcal{L}_{\mathrm{\Theta}^{+}},
\end{eqnarray}
\begin{eqnarray}
{\mathcal{L}_{\mathrm{N}}}=  \bar{\Psi}_{N} (
i\gamma^{\mu}\partial_{\mu}- M_{N})\Psi_{N}  -
g_{\sigma}^{N}\bar{\Psi }_{N} \sigma\Psi_{N}-
 g_{\omega}^{N}\bar{\Psi}_{N}\gamma^{\mu}\omega_{\mu}\Psi_{N}\cr- g_{\rho}^{N}
 \bar{\psi}_{N} \gamma^{\mu}\rho^{a}_{\mu}\tau^{a}\Psi_{N}
  +\frac{1}{2}\partial^{\mu}\sigma\partial_{\mu}\sigma-\frac{1}{2}m
_{\sigma}^{2}\sigma^{2}- \frac{1}{3}g
_{2}^{2}\sigma^{3}\cr-\frac{1}{4} g _{3}^{2}\sigma^{4}
 -\frac{1}{4}\Omega^{\mu\nu}\Omega_{\mu\nu} +\frac{1}{2}
m_{\omega}^{2}\omega^{\mu}\omega_{\mu} -\frac{1}{4}  R ^{a \mu
\nu}   R _{\mu\nu}^{a}\cr+ \frac{1}{2} m_{\rho}^{2} {\rho^{a\mu}}
{\rho ^{a}_{\mu}} -\frac{1}{4} F^{\mu\nu}F_{\mu\nu}-e \bar{\psi
}_{N}\gamma^{\mu} A^{\mu}\frac{1}{2}(1-\tau^{3})\psi_{N}.
\end{eqnarray}
\begin{eqnarray}
{\mathcal{L}_{\mathrm{\Theta}^{+}}} = \bar{\Psi}_{{\Theta}^{+}} (
i\gamma^{\mu}\partial_{\mu}- M_{{\Theta}^{+}})\Psi_{{\Theta}^{+}}-
g_{\sigma}^{ \Theta } \bar{\Psi}_{{\Theta}^{+}}
\sigma\Psi_{{\Theta}^{+}}-
 g_{\omega}^{ \Theta }\bar{\Psi}_{\Theta^{+}}\gamma^{\mu}\omega_{\mu}\Psi_{{\Theta}^{+}}\cr- g_{\rho}^{ \Theta }
\bar{ \Psi}_{{\Theta}^{+}}
\gamma^{\mu}\rho^{a}_{\mu}\tau^{a}\Psi_{{\Theta}^{+}}+e \bar{\Psi
}_{{\Theta}^{+}}\gamma^{\mu}
A^{\mu}\frac{1}{2}(1-\tau^{3})\Psi_{{\Theta}^{+}}
\end{eqnarray}
with
\begin{eqnarray}
 \Omega^{\mu\nu}&=&\partial^{\mu}\omega^{\nu}-\partial^{\nu}\omega^{\mu},\cr
 R ^{a\mu\nu}&=&\partial^{\mu}\rho^{a\nu}-\partial^{\nu}\rho^{a\mu},\cr
 F ^{\mu\nu}&=&\partial^{\mu}A^{\nu}-\partial^{\nu}A^{\mu}.
\end{eqnarray}

where the meson fields are denoted by $\sigma$, $\omega_{\mu}$
$\rho_{\mu}$, and their masses by $m_{\sigma}$, $m_{\omega }$,
$m_{\rho}$, respectively;  $\Psi_{N}$ and $\Psi_{{\Theta}^{+}}$
are the nucleon and ${\Theta}^{+}$ baryon fields with
corresponding masses $M_{N}$ and $M_{{\Theta}^{+}}$, respectively;
$A_{\mu}$ is the electromagnetic field.
$g_{\sigma}^{N}$($g_{\sigma}^{\Theta}$),
$g_{\omega}^{N}$($g_{\omega}^{\Theta}$) and
$g_{\rho}^{N}$($g_{\rho}^{\Theta}$) are the
$\sigma-N$($\sigma-\Theta$), $\omega-N$($\omega-\Theta$) and
$\rho-N$($\rho-\Theta$) coupling constants , respectively. The
isospin Pauli matrices are written as $\tau^{a}$, $\tau^{3}$ being
the third component of $\tau^{a}$. For ${\Theta}^{+}$ baryon, we
assume isospin $I=0$, $\tau^{a}=0$, there is no coupling to $\rho$
meson.
\section{How to determine the coupling constants}

   According the quark meson coupling model developed in Ref.\cite{f}, the baryon coupling constants can be
   determined. The equation of motion for meson field operators
   are
   \begin{eqnarray}
   \partial_{\mu}\partial ^{\mu}
   \hat{\sigma}+m_{\sigma}^{2}\hat{\sigma}=g_{\sigma}^{q}\bar{q}q,\\
   \partial_{\mu}\partial ^{\mu}
   \hat{\omega}^{\nu}+m_{\omega}^{2}\hat{\omega}^{\nu}=g_{\omega}^{q}\bar{q}\gamma^{\nu}q,\\
   \partial_{\mu}\partial ^{\mu}
   \hat{\rho}^{\nu,\alpha}+m_{\rho}^{2}\hat{\rho}^{\nu,\alpha}
   =g_{\rho}^{q}\bar{q}\gamma^{\nu}\frac{\tau^{\alpha}}{2}q,
   \end{eqnarray}
where $g_{\sigma}^{q}$, $g_{\omega}^{q}$ and $g_{\rho}^{q}$ are
the quark-meson coupling constants  for $\sigma$, $\omega$ and
$\rho$, respectively.

The mean fields are defined as the expectation values in the
ground state of the nucleus, $|A\rangle$:
\begin{eqnarray}
\langle A|\hat{\sigma}(t,r)|A \rangle=\sigma(r),\\
\langle A|\hat{\omega}^{\nu}(t,r)|A \rangle=\delta(\nu,0)\omega(r),\\
\langle A|\hat{\rho}^{\nu,\alpha}(t,r)|A
\rangle=\delta(\nu,0)\delta(\alpha,3)\rho(r),
\end{eqnarray}
where (t,r) are the coordinates in the rest frame of the nucleus.
   In the mean field approximation the sources are the sum of the sources
   created by each nucleon - the latter interacting with the meson
   fields. Thus we write
   \begin{eqnarray}
 \bar{ q}q(t,r)= \sum_{i=1, A}\langle \bar{ q}q(t,r)
 \rangle_{i},
  \\
  \bar{q}\gamma^{\nu}q(t,r)= \sum_{i=1, A}\langle  \bar{q}\gamma^{\nu}q(t,r)
 \rangle_{i},
 \end{eqnarray}
 where $\langle...\rangle_{i}$ denotes the matrix element in the
 nucleon $i$ located at $\mathbf{\emph{R}}_{i}$ at time $t$.
 According to the Born-Oppenheimer approximation, the baryon
 structure is described by $n$ quarks in the lowest mode.
 therefore, in the instantaneous rest frame (IRF) of the nucleon
 $i$, we have
 \begin{eqnarray}
\langle \bar{q} 'q'(t',r')\rangle_{i}= n\sum_{m}
\bar{\phi}^{0,m}_{i}(u')\phi^{0,m}_{i}(u')=ns_{i}(u')
  ,
  \\
 \langle \bar{q} '\gamma^{\nu}q'(t',r')\rangle_{i}=
 n\delta(\nu,0)\sum_{m} \bar{\phi} ^{\dag
 0,m}_{i}(u')\phi^{0,m}_{i}(u')=n\delta(\nu,0)w_{i}(u'),
 \end{eqnarray}
 where $(t',r')$ are the the coordinates  and  $q'$ is the quark
 field in the IRF. $\phi^{0,m}_{i}$ is the complete and orthogonal
 set of eigenfunctions which are defined by Eq.(23) in
 Ref.\cite{f}.

 From the Lorents transformation properties of the fields we get
 \begin{eqnarray}
  \langle \bar{q}q(t,r)\rangle_{i}=\frac{n}{(2\pi)^{3}}((\cosh\xi_{i}))^{-1}\int d k  e^{ik\cdot(r-R_{i}))} S(k,R_{i}),
  \\
 \langle \bar{q}\gamma^{0}q(t,r)\rangle_{i}=
 \frac{n}{(2\pi)^{3}}\int d k  e^{ik\cdot(r-R_{i}))} W(k,R_{i}),
 \\
\langle \bar{q}  \gamma q(t,r)\rangle_{i}=
 \frac{n}{(2\pi)^{3}}v_{i}\int d k  e^{ik\cdot(r-R_{i}))}
 W(k,R_{i}),
 \end{eqnarray}
 with
 \begin{eqnarray}
S(k,R_{i})=\int du e^{-i(k_{\perp}\cdot
u_{\perp}+k_{L}u_{L}/\cosh\xi_{i})}s_{i}(u),
\\
W(k,R_{i})=\int du e^{-i(k_{\perp}\cdot
u_{\perp}+k_{L}u_{L}/\cosh\xi_{i})}w_{i}(u).
\end{eqnarray}
The corresponding physical quantities such as $\xi_{i}$,
$u_{\perp}$, $u_{L}$ etc. are defined in Ref.\cite{f}. Finally,
the mean field expressions for the meson sources take the form
\begin{eqnarray}
 \langle A \mid\overline{q}q(t,r)\mid A \rangle =\frac{n}{(2\pi)^{3}}
 \int d k e^{ik\cdot r} \langle A\mid\sum_{i}((\cosh\xi_{i}))^{-1} e^{-ik\cdot R_{i} } S(k,R_{i})\mid A\rangle,
\\
\langle A \mid\overline{q}\gamma^{0}q(t,r)\mid A \rangle
=\frac{n}{(2\pi)^{3}}\int d k e^{ik\cdot r} \langle A\mid\sum_{i}
e^{-ik\cdot R_{i} } W(k,R_{i})\mid A\rangle,
\\
\langle A \mid\overline{q}\gamma q(t,r)\mid A \rangle=0.
\end{eqnarray}
To simplify further, we remark that a matrix element of the form
\begin{equation}
\langle A\mid\sum_{i} e^{-ik\cdot R_{i} } ...\mid A\rangle
\end{equation}
is negligible unless $k$ is less than, or of the order of, the
reciprocal of the nuclear radius.

We now define the scalar, baryon and isospin densities of the
nucleons in the nucleus by
\begin{eqnarray}
\rho_{s}=\langle A\mid\sum_{i}
\frac{M_{B}^{*}(R_{i})}{E_{i}-V(R_{i})}\delta(r-R_{i})\mid
A\rangle,
\\
\rho_{B}=\langle A\mid\sum_{i}
 \delta(r-R_{i})\mid
A\rangle,
\\
\rho_{3}=\langle A\mid\sum_{i}\frac{\tau_{3}^{B}}{2}
 \delta(r-R_{i})\mid
A\rangle,
\end{eqnarray}
where we have used $E_{i}=M_{B}^{*}\cosh \xi_{i}+V(R_{i})$
(Eq.(58) of Ref.\cite{f}) to eliminate the factor $(\cosh
\xi_{i})^{-1}$. The meson sources take the simple form
\begin{eqnarray}
 \langle A \mid \bar{q}q(t,r)\mid A \rangle =nS(r)\rho_{s},
\\
\langle A \mid \bar{q} \gamma^{0}q(t,r)\mid A \rangle =n\rho_{B},
\\
\langle A \mid \bar{q} \gamma^{\nu}\frac{\tau
^{\alpha}}{2}q(t,r)\mid A
\rangle=\delta(\nu,0)\delta(\alpha,3)\rho(3).
\end{eqnarray}
We have used the notation

\begin{eqnarray}
S(r)=S(0,r)&=&\int d u s_{r}(u) =
\frac{\Omega_{0}/2+m_{q}^{*}R_{B}(\Omega_{0}-1)}{\Omega_{0}(\Omega_{0}-1)+m_{q}^{*}R_{B}/2},\cr
\Omega_{0}&=&\sqrt{\chi^{2}+(R_{B}m_{q}^{*})^{2}},\cr
m_{q}^{*}&=&m_{q}-g_{\sigma}^{q}\sigma(r)
\end{eqnarray}
Here, $\chi$ and $m_{q}$ are the parameters for lowest eigenvalues
for the quarks and the corresponding current quark masses. $R_{B}$
is the radius of the bag.

Since their sources are time independent and since they do not
propagate, the mean fields are also time independent. So, we get
the desired equations for $\sigma(r)$, $\omega(r)$ and $\rho(r)$:

\begin{eqnarray}
   \{-\Delta+m_{\sigma}^{2}\} \sigma(r)=
   g_{\sigma}C(r)\rho_{s}(r),
\\
   \{-\Delta+m_{\omega}^{2}\}\omega_{0}(r)=g_{\omega}\rho_{B}(r),
 \\
   \{-\Delta+m_{\rho}^{2}\} \rho_{00}(r)=g_{\rho}\rho_{3}(r).
\end{eqnarray}
where the nucleon coupling constants and $C(r)$ are defined by
\begin{eqnarray}
g_{\sigma}=ng_{\sigma}^{q}S(\sigma=0),
   g_{\omega}=ng_{\omega}^{q}, g_{\rho}=g_{\rho}^{q},
   C(r)=S(r)/S(\sigma=0).
\end{eqnarray}

In an approximation where the $\sigma(r)$ $\omega(r)$ and
$\rho(r)$ fields couple only to the $u$ and $d$
quarks\cite{f1,f2}, the coupling constants in the strange baryons
are obtained as $g_{\omega}^{S}=(n_{q}/n)g_{\omega}$ and
$g_{\rho}^{S}\equiv g_{\rho}=g_{\rho}^{q}$, with $n_{q}$ being the
total number of valence $u$ and $d$ quarks in the baryon S. For
the strange baryon S, the coupling constant $\sigma-S$ can be
written as
\begin{eqnarray}
g_{\sigma}^{S}=n_{q}g_{\sigma}^{q}S_{S}(\sigma=0),
\end{eqnarray}
where
\begin{equation}
S_{S}(r)=\frac{\Omega_{0}/2+m_{q}^{*}R_{S}(\Omega_{0}-1)}{\Omega_{0}(\Omega_{0}-1)+m_{q}^{*}R_{S}/2},
\end{equation}
where $R_{S}$ is the bag radius for the strange baryon. By
combining the Eq.(24) and Eq.(25) we get
\begin{equation}
g_{\sigma}^{S}=\frac{n_{q}}{n}g_{\sigma}S_{S}(\sigma=0)/S(\sigma=0)=\frac{n_{q}}{n}g_{\sigma}\Gamma_{S/B}.
\end{equation}
According Eq.(24)
\begin{eqnarray}
g_{\sigma}=\frac{n_{q}}{3}g_{\sigma}^{N} ,
   g_{\omega}=\frac{n_{q}}{3}g_{\omega}^{N}, g_{\rho}=g_{\rho}^{N},
\end{eqnarray}
where $g_{\sigma}^{N}$, $g_{\omega}^{N}$, $g_{\rho}^{N}$ are the
coupling constants for nucleon. So, we can easily get the
relations
\begin{eqnarray}
g_{\sigma}^{S}=\frac{n_{q}}{3}g_{\sigma}^{N}\Gamma_{S/B} ,
   g_{\omega}^{S}=\frac{n_{q}}{3}g_{\omega}^{N},
   g_{\rho}^{S}=g_{\rho}^{N}.
\end{eqnarray}

 \section{Calculations and Analysis}
In this section we will show the our numerical results using
relativistic mean field theory model. In the mean-field and the
no-sea approximations\cite{aa} the contributions of anti-(quasi)
particles and quantum fluctuations of mesons fields can be thus
neglected. The contribution of the tensor coupling is neglected
for simple, although it is considered in some Refs.\cite{e1,e2}.
The coupling constants of nucleon to mesons employed are
NL-SH\cite{g}, which are list in table 1. The coupling constants
of $\Theta^{+}$ to mesons are obtained from the relations Eq.(39).
For $\Theta^{+}$, $n_{q}$=4, isospin $I=0$ means no coupling with
$\rho$ meson. Thus, we have the relations
\begin{eqnarray}
g_{\sigma}^{\Theta}=\frac{4}{3}g_{\sigma}^{N}\Gamma_{\Theta/B} ,
   g_{\omega}^{\Theta}=\frac{4}{3}g_{\omega}^{N}.
\end{eqnarray}
In practice the value for $\Gamma_{S/B}\approx1$  for all hyperons
even though $R_{B}\neq R_{S}$\cite{aa1}, thus in our calculation,
we use $\Gamma_{\Theta/B}=1$.

\subsection{The results for $\Theta^{+}$ hypernuclei in RMF }
In figure 1, we show the single-particle energy spectrum such as
1s1/2, 1p3/2, 1p1/2, 1d5/2... of $\Theta^{+}$ in $^{6}{\rm
Li}$,$^{12}{\rm C}$, $^{16}{\rm O}$, $^{40}{\rm Ca}$ and
$^{208}{\rm Pb}$, which is denoted by 1, 2a, 3, 4a, 5
respectively. 2b is the energy spectrum for $^{12}_{\Theta}{\rm
C}$ when the coupling constant change a little to
$g_{\sigma}^{\Theta}=\frac{4}{3}g_{\sigma}^{N}\times0.95$. 2c, 2d
and 4b, 4c are the results for $^{13}_{\Theta}{\rm C}$ and
$^{41}_{\Theta}{\rm Ca}$ at the depth of potential 60 MeV and 120
MeV respectively of ref\cite{c2}.

In light nuclei like $^{6}{\rm Li}$ (1), $^{12}{\rm C}$(2a) and
$^{16}{\rm O}$(3) we find several bound states. The separation of
the two deepest bound states (1s1/2 and 1p3/2) is about 24 MeV for
$^{6}{\rm Li}$, about 60 MeV for $^{12}{\rm C}$ and around 30 MeV
for $^{16} {\rm O}$. For medium and heavy nuclei, as in $^{40}{\rm
Ca}$,$^{208}{\rm Pb}$ shown in the figure, we find more bound
states and the energy separation between 1s1/2 and 1p3/2 is about
18 MeV for $^{40}{\rm Ca}$ and about 4 MeV for $^{208}{\rm Pb}$.
There is a trend that the separation between 1s1/2 and 1p3/2
becomes smaller and smaller and the bound state becomes more and
more with the increasing of the nucleon number. It is interesting
that the width of $\Theta^{+}$ is very narrow with upper limit as
small as 15 MeV predicted by the chiral soliton model in Ref.
\cite{a}, and 9 MeV\cite{h}.Studies based on $K^{+}N$ scattering
suggest that the width should be smaller than 5
MeV\cite{h2,h3,h4}. Cahn and trilling have extracted
$\Gamma(\Theta)=0.9\pm0.3$ MeV from an analysis of Xenon bubble
chamber\cite{h1}. The narrow upper limit width of $\Theta^{+}$ is
often smaller than the separation of the two deepest bound states
(1s1/2 and 1p3/2) for light and medium nuclei, which would make a
clear case for the experimental observation of these states.

The splitting of spin-orbit for $\Theta^{+}$ hypernuclei can be
get from our calculation with RMF. Such as, from figure 1, we can
find that the splitting between 1p3/2 and 1p1/2 is about 5 MeV for
$^{6}{\rm Li}$ (1) and $^{16}{\rm O}$ , 10 MeV for $^{12}{\rm C}$,
The splitting is smaller for medium nucleus $^{40}{\rm Ca}$ (about
2 MeV) and there is no obvious splitting for heavy nucleus
$^{208}{\rm Pb}$. If the prediction based on $K^{+}N$ scattering
suggest that the width should be smaller than 5 MeV\cite{h2,h3,h4}
is true, the the splitting between 1p3/2 and 1p1/2 is also bigger
than the narrow upper limit width of $\Theta^{+}$, which means
that we may observe splitting of spin-orbit for light $\Theta^{+}$
hypernuclei. There is a very interesting phenomenon that both
separation between 1s1/2 and 1p3/2 and splitting of spin-orbit
between 1p3/2 and 1p1/2 are much larger than the others.

For Comparison, we also yield the the single-particle energy
spectrum of $\Theta^{+}$ from Ref\cite{c2} by solving
Schr\"{o}dinger equation with two potentials. It is a very simple
model and can not give the splitting of spin-orbit. It estimates
that the binding energy(1s) of $^{12}{\rm C}$ is in the region
$34.0\sim87.3$ MeV and $42.6\sim98.2$ MeV for $^{40}{\rm Ca}$. In
our calculation, the binding energy(1s) of $^{12}{\rm C}$ is about
116 MeV and 81 MeV for $^{40}{\rm Ca}$, and the energy separation
between two bound states is about 18 MeV for $^{40}{\rm Ca}$ and
60 MeV for $^{12}{\rm C}$. Our results are very close to the the
results estimated by solving Schr\"{o}dinger equation in
ref\cite{c2}.

In table 1 we list the calculated binding energy per baryon,
$-E/A$, r.m.s. charge radius, $r_{ch}$, and r.m.s. radii of the
$\Theta^{+}$ baryon and the neutron and proton distributions ($r_{\Theta}$ , $r_{n}$ and $r_{p}$).
For comparision, we also give these quantities without $\Theta^{+}$.
The differences in vulues for finite nuclei and hypernuclei listed in table 1
reflects the effects of the $\Theta^{+}$. The appearence of $\Theta^{+}$ in nuclei leads to
larger binding energy per baryon, $-E/A$. From table 1, it is found that
 the values of   $r_{n}$, $r_{p}$ and $r_{ch}$  for
 $^{9}_{\Theta} {\rm Be}$, $^{13}_{\Theta}{\rm C}$
are obvious less than those for the corresponding normal nuclei.
This is the shrinkage effect which has been found in lighter
$\Lambda$ hypernuclei\cite{e3,g1}. In our calculation for
$^{9}_{\Theta}{\rm Be}$, $^{13}_{\Theta}{\rm C}$, we also find
that there is shrinkage effect and  gluelike role for
$\Theta^{+}$. But this phenomenon is not found for
$^{7}_{\Theta}{\rm Li}$ and it is not obvious for
$^{17}_{\Theta}{\rm O}$, $^{41}_{\Theta}{\rm Ca}$ and
$^{209}_{\Theta}{\rm Pb}$.

The key point for our calculation is how to determine the coupling
constants of $\Theta^{+}$ to mesons as accurate as possible. In
the calculation, we find that the binding energy is very sensitive
to the coupling constants. In Refs\cite{aa2}, they determine the
the coupling constants of hyperon to mesons by adjusting
$g_{\sigma}^{Y}/g_{\sigma}^{N}$ around 2/3 (according
Eq.(39),$g_{\sigma}^{Y}/g_{\sigma}^{N}\approx2/3$ )
  to fit the experimental data
because of the effect of medium in nuclei.Where $Y$ is stands the
hyperon. To consider the effect of medium to the coupling constant
$g_{\sigma}^{\Theta}$, we list the result of $^{12}{\rm C}$ in
figure 1, denoted with 2b, when the coupling constant
$g_{\sigma}^{\Theta}$ changes to $g_{\sigma}^{\Theta}\times$0.95.
From the figure, we see the binding energy of 1s1/2  have a large
change from 116 to 60 MeV, but there is still  several bound
states and the energy separation energy between 1s1/2 and 1p3/2 is
around 32 MeV.

We also the list results of the ground state for $^{12}{\rm C}$
when the coupling constant $g_{\sigma}^{\Theta}$ changes from
13.925 to 11.725 ($g_{\sigma}^{\Theta}\times$0.84) in table 2.
From it, we can see that the binding energy of 1s1/2 is sensitive
to the coupling constant, if
$g_{\sigma}^{\Theta}<11.725(13.925\times0.84)$, there are no bound
states for $\Theta^{+}$ in $^{12}{\rm C}$. In Ref.\cite{aa2},
$g_{\sigma}^{Y}/g_{\sigma}^{N}$ is 0.621 ($0.667\times0.93$) and
0.619 ($0.667\times0.928$) for $\Lambda-\sigma$ and
$\Sigma-\sigma$, respectively. Thus, from the calculation of
before\cite{aa2},we can conclude that the effect of medium is not
very large, the considered medium effective coupling constant is
about $0.93\times g_{\sigma}^{Y}$ for hyperon,  If we let
$g_{\sigma}^{\Theta}=0.93\times13.925$=12.95, from table 1, we can
see that the binding energy of 1s1/2 for $^{12}{\rm C}$ is between
52 and 40 MeV, which is agree with the estimation $34\sim 87$ MeV
in the Ref\cite{c2}.

In table 2, the binding energy per baryon, -E/A (in MeV), r.m.s.
charge radius $r_{ch}$ (in fm), r.m.s. radii of
 $\Theta^{+}$, neutron and proton, $r_{\Theta}$ , $r_{n}$ and $r_{p}$ (in fm)
 are also shown respectively. It is seen in table 2 that the radii is
 increasing and the binding energy per baryon is decreasing with the
 decreasing of the coupling constant. When
 $g_{\sigma}^{\Theta}<12.325$, $r_{\Theta}<r_{ch}$, which means
 that the $\Theta^{+}$ is out of the nucleus $^{12}{\rm C}$. It can only
 form $\Theta^{+}$ atom.

\begin{table}[h]
\begin{center}
\caption{\footnotesize The parametrization of the nucleonic sector
(adopted from Ref.\cite{g}). The masses are given in MeV and the
coupling $g_{2}$ in fm$^{-1}$. } \label{parameter} \vspace{0.1cm}
\begin{tabular}{|c|c|c|c|c|c|}\hline\hline
\ \ $M_{N}$\ \ & \ \ \ 939.0  \ \ & \ \  $g_{\sigma }^{N}$ \ \ \ & \ \ \ 10.444 \ \ \ \ &$g_{3}$& \ \ \ -15.8337 \ \ \ \\
$m_{\sigma}$ & 526.059 & $g_{ \omega}^{N}$& 12.945 &  & \\
$m_{\omega}$ & 783.0 & $g_{ \rho}^{N}$ & 4.383 & & \\
$m_{\rho}$ & 763.0 &  $g_{2}$  &  -6.9099   &&\\
 \hline
 \end{tabular}
 \end{center}
 \end{table}

\begin{table}[h]
\begin{center}
\caption{ Binding energy per baryon, $-E/A$ (in MeV), binding
energy of 1s1/2 for $\Theta^{+}$ in nuclei, $-B$ (in MeV), r.m.s.
charge radius $r_{ch}$ (in fm), r.m.s. radii of hyperon, neutron
and proton,$r_{\Theta}$ , $r_{n}$ and $r_{p}$ (in fm),
respectively. The configuration of $\Theta^{+}$ is 1s1/2 for all
hypernclei } \label{parameter} \vspace{0.1cm}
\begin{tabular}{|c|c|c|c|c|c|c|c |}\hline\hline
$^{A}Z$&$r_{p}$(fm)&$r_{n}$(fm)&$r_{\Theta}$(fm)&$r_{ch}$(fm)&$-E/A$(MeV)&$-B$(MeV)\\
\hline
$^{6}$Li           & 2.37  & 2.32  &     & 2.51 & 5.67  & \\
$^{7}_{\Theta}$Li  & 2.36  & 2.36  &1.38 & 2.50 & 6.21 &52.13\\
\hline
$^{8}$Be           & 2.34  & 2.30  &     & 2.48 & 5.42 &  \\
$^{9}_{\Theta}$Be  & 2.21  & 2.19  &1.35 & 2.36 & 11.43 &67.72 \\
\hline
$^{12}$C           & 2.32  & 2.30  &     & 2.46 & 7.47 &  \\
$^{13}_{\Theta}$C  & 2.17  & 2.15  &1.19 & 2.32 & 14.41  &116.24 \\
\hline
$^{16}$O           & 2.58  & 2.55  &     & 2.70 & 8.04  &  \\
$^{17}_{\Theta}$O  & 2.58  & 2.54  &1.46 & 2.70 & 8.60  &83.66 \\
\hline
$^{40}$Ca          & 3.36  & 3.31  &     & 3.46 & 8.52  &  \\
$^{41}_{\Theta}$Ca & 3.35  & 3.30  &1.86 & 3.45 & 10.18  &81.35  \\
\hline
$^{208}$Pb         & 5.45  & 5.71  &     & 5.51 & 7.90  &  \\
$^{209}_{\Theta}$Pb & 5.44  & 7.70 &3.87 & 2.49 & 8.19  &69.54  \\
\hline
\end{tabular}
\end{center}
 \end{table}

\begin{table}[h]
\begin{center}
\caption{ Binding energy per baryon, $-E/A$ (in MeV), binding
energy of 1s1/2 for $\Theta^{+}$ in nuclei, $-B$ (in MeV), r.m.s.
charge radius $r_{ch}$ (in fm), r.m.s. radii of hyperon, neutron
and proton,$r_{\Theta}$ , $r_{n}$ and $r_{p}$ (in fm),
respectively. The configuration of $\Theta^{+}$ is 1s1/2 for
$^{13}_{\Theta}{\rm C}$ in different $g_{\sigma}^{\Theta}$. }
\label{rr} \vspace{0.1cm}
\begin{tabular}{|c|c|c|c|c|c|c|c |}\hline\hline
$g_{\sigma}^{\Theta}$&$r_{p}$(fm)&$r_{n}$(fm)&$r_{\Theta}$(fm)&$r_{ch}$(fm)&$-E/A$(MeV)&$-B$(MeV)\\
\hline
13.925 & 2.17  & 2.15  &1.19 & 2.32 & 14.24 &116.24\\
13.725 & 2.18  & 2.16  &1.22 & 2.33 & 13.27 &100.00\\
13.525 & 2.19  & 2.17  &1.28 & 2.34 & 12.33 &83.70 \\
13.325 & 2.21  & 2.19  &1.35 & 2.36 & 11.43 &67.72 \\
13.125 & 2.24  & 2.21  &1.46 & 2.38 & 10.60 &52.92 \\
12.925 & 2.26  & 2.24  &1.59 & 2.41 & 9.85  &40.17 \\
12.725 & 2.29  & 2.27  &1.76 & 2.44 & 9.19  &29.67 \\
12.525 & 2.32  & 2.29  &1.95 & 2.46 & 8.60  &21.20 \\
12.325 & 2.34  & 2.31  &2.20 & 2.48 & 8.10  &14.43 \\
12.125 & 2.35  & 2.32  &2.50 & 2.49 & 7.68  &9.08  \\
11.925 & 2.35  & 2.32  &2.90 & 2.49 & 7.34  &4.98  \\
11.725 & 2.35  & 2.32  &3.43 & 2.49 & 7.10  &2.01  \\
\hline
\end{tabular}
\end{center}
 \end{table}

\subsection{$\Theta^{+}$ potential depth  in nuclear matter}

In nuclear matter the $\Theta^{+}$ potential depth is written
as\cite{e3}
\begin{eqnarray}
U=g_{\sigma}^{\Theta}\sigma^{eq}+g_{\omega}^{\Theta}\omega_{0}^{eq}
\end{eqnarray}
where $\sigma^{eq}$ and $\omega_{0}^{eq}$  are the values of
$\sigma$ and $\omega$ fields at the saturation nuclear density,
respectively. Use the relations
$\sigma^{eq}=(M^{*}_{N}-M_{N})/g_{\sigma}^{N}$ and
$\omega_{0}^{eq}=g_{\omega}^{N}\varrho_{0}/m_{\omega}^{2}$ at the
saturation nuclear density, we can get\cite{g2}
\begin{eqnarray}
U=M_{N}(\frac{M^{*}_{N}}{M_{N}}-1)\frac{g_{\sigma}^{\Theta}}{g_{\sigma}^{N}}
+(\frac{g_{\omega}^{N} }{m_{\omega}})^{2}
\frac{g_{\omega}^{\Theta}}{g_{\omega}^{N}}\varrho_{0}
\end{eqnarray}
where $M^{*}_{N}$ is the effective mass of nucleon and
$\varrho_{0}$ is the saturation nuclear density. From Eq.(42), we
can see that the depth of potential $U$ is determined by
$\frac{M^{*}_{N}}{M_{N}}$,
$\frac{g_{\sigma}^{\Theta}}{g_{\sigma}^{N}}$ and $\varrho_{0}$, if
we let $\frac{g_{\omega}^{\Theta}}{g_{\omega}^{N}}=4/3$ as in
Eq.(40). Usually, the effective mass $\frac{M^{*}_{N}}{M_{N}}$
 and saturation nuclear density $\varrho_{0}$ are considered as
 "experimental input". In the RMF calculation, $\frac{M^{*}_{N}}{M_{N}}\sim
 0.60$, $\varrho_{0}\sim 0.15$fm$^{-3}$\cite{g3}. Here, we let them have
 some uncertain and can change in a little region. According the
 calculation and analysis of section 3, when $g_{\sigma}^{\Theta}<11.725(13.925\times0.84)$,
there are no bound states for $\Theta^{+}$ in $^{12}{\rm C}$, thus
we let it as a approximation that there is no bound states for
$\Theta^{+}$ in nuclear matter when
$g_{\sigma}^{\Theta}<11.725(13.925\times0.84)$, namely the depth
of potential $U\geq0$  .

 In figure 2, we show the relation of the potential depth changing with
 the effective mass $\frac{M^{*}_{N}}{M_{N}}$ for  $\varrho_{0}=
 0.15$,$\varrho_{0}=0.16$ and $\varrho_{0}=0.17$fm$^{-3}$ in a,b
 and c respectively. From figure 2(a), it is easily seen that
 $U$=0 corresponds to effective mass $\frac{M^{*}_{N}}{M_{N}}=0.60$ when
$\varrho_{0}=0.15$fm$^{-3}$. With the same method we can determine
 $\frac{M^{*}_{N}}{M_{N}}=0.575$ for $\varrho_{0}=
 0.16$fm$^{-3}$ and $\frac{M^{*}_{N}}{M_{N}}=0.546$ for $\varrho_{0}=
 0.17$fm$^{-3}$ from b and c. The results of  effective mass
 $\frac{M^{*}_{N}}{M_{N}}$ is reasonable, because they are around
 0.6, which agrees with the fit data in RMF\cite{g3}.

  Since the effective mass
 $\frac{M^{*}_{N}}{M_{N}}$ and saturation nuclear density
 $\varrho_{0}$ are determined, we can get the potential depth at
 saturation nuclear density easily. From figure 2 (a), we can see $U$=-81 MeV
 at $g_{\sigma}^{\Theta}=13.925$ and $U$=-46 MeV at
 $g_{\sigma}^{\Theta}=13.925\times0.93$ when
 $\frac{M^{*}_{N}}{M_{N}}=0.60$,$\varrho_{0}=0.15$fm$^{-3}$.
 Considered the effect of medium, if the $\sigma-\Theta^{+}$ coupling
 constant is in the region $13.925\leq g_{\sigma}^{\Theta}\leq
 13.925\times 0.93$, the potential depth is in the region $-81 \leq U \leq
 -46$ MeV. From figure 2(b) and (c), we also get $-84 \leq U \leq
 -47$ MeV for $\varrho_{0}=0.16$fm$^{-3}$ and $-92 \leq U \leq
 -52$ MeV for $\varrho_{0}=0.17$fm$^{-3}$ when $13.925\leq g_{\sigma}^{\Theta}\leq
 13.925\times 0.93$ in the same way. From the calculation, it is
 found that the uncertain of the saturation nuclear density can
 bring some uncertain for the potential depth in several MeV.
 Because there is still no experimental data for $\Theta^{+}$ in nuclear
 matter, we can only estimate the potential depth in theory, the
 uncertain for the potential depth in several MeV is allowed. In
 fact, the potential depth estimated by us agrees well with that
 of in Ref.\cite{c2}$-120 \leq U \leq
 -60$ MeV.

\section{Summary}
We have educed the coupling constants for strange many quarks
system with quark meson coupling model. We get the relation
$g_{\sigma}^{\Theta}=\frac{4}{3}g_{\sigma}^{N}\Gamma_{\Theta/B}$ ,
$g_{\omega}^{\Theta}=\frac{4}{3}g_{\omega}^{N}$ for $\Theta^{+}$
coupling with $\sigma$ and $\omega$ meson. With the coupling
constants for $\sigma-\Theta^{+}$ and $\omega-\Theta^{+}$ educed
by us, the calculations for $\Theta^{+}$ in nuclei are carried out
in the framework of RMF. From the single-particle levels of
$\Theta^{+}$ in nuclei, it is found that there are several bound
states for light nuclei, such as $^{7}_{\Theta}{\rm Li}$, and more
bound states for heavier nuclei. The binding energies of 1s1/2 are
usually as high as 50 MeV to 120 MeV, which means that there are
strong attractive interaction for $\Theta^{+}$-nucleus. This
prediction of RMF is agree with the prediction of
Refs.\cite{c1,c2}. The strong attractive interaction for
$\Theta^{+}$-nucleus induces stronger per nucleon binding energy
than normal nuclei, which can be seen obviously in table 2.
Because of the strong attractive interaction for
$\Theta^{+}$-nucleus, the shrinking effect is found for light
nuclei $^{9}_{\Theta}{\rm Be}$ and $^{13}_{\Theta}{\rm C}$, in
medium or heavy nuclei there is no obvious shrinking effect. The
shrinking effect is first found in the light
$\Lambda$-hypernucleus  for lithium hypernuclei\cite{g}, but there
is no shrinking effect for $^{7}_{\Theta}{\rm Li}$ in RMF
prediction.

The depth of optical potential for $\Theta^{+}$ in nuclear matter
is investigated, the region is given. If we assume the saturation
nuclear density $\varrho_{0}=0.15$ 0.16 and 0.17 fm$^{-3}$ with
the coupling constant $13.925\leq g_{\sigma}^{\Theta}\leq
 13.925\times 0.93$, the
depth of potential $-81 \leq U \leq -46$ MeV, $-84 \leq U \leq
 -47$ MeV and $-92 \leq U \leq -52$ MeV, respectively.

 In this work we assume that $J^{P} = 1/2^{+}$ and $I=0$ for
 $\Theta^{+}$, and when we induce the coupling constants in
 section 3 with QMC model, we do an approximation where the $\sigma(r)$ $\omega(r)$ and
$\rho(r)$ fields couple only to the $u$ and $d$ quarks. These
assume and approximation may be affect our results.

 \hspace{5cm}{ACKNOWLEDGEMENTS}

This work was supported by National Natural Science Foundation of
China (10275037) and Specialized Research Fund for the Doctoral
Program of Higher Education of China (20010055012).

\begin{figure}[ht]
\centering\includegraphics[width=16.cm,height=12cm]{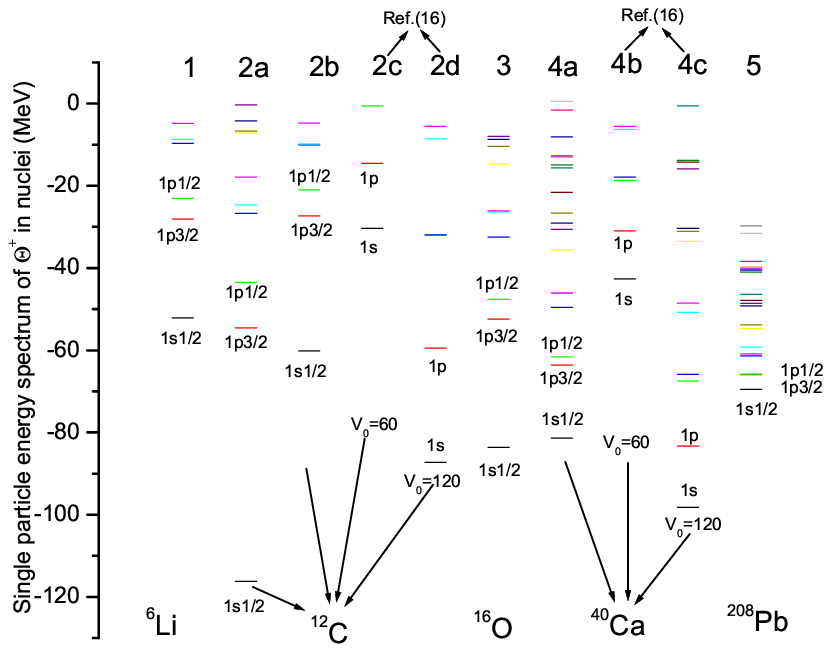}
%\epsfbox{graph4.eps}
\caption{ Single-particle levels for $\Theta^{+}$ in $^{6}{\rm
Li}$, $^{12}{\rm C}$, $^{16}{\rm O}$, $^{40}{\rm Ca}$ and
$^{208}{\rm Pb}$ denoted with 1,2a,2b,2c,2d,3,4a,4b,4c and 5
respectively. 1, 2a, 3, 4a, 5 are our results with RMF at
$g_{\sigma}^{\Theta}=13.925$ and 2b is our result at
$g_{\sigma}^{\Theta}=13.925\times0.95$. 2c,2d and 4b,4c are the
results from Ref.\cite{c2} in $^{12}{\rm C}$ and $^{40}{\rm Ca}$
at the depth of potential 60 MeV and 120 MeV respectively.}
\end{figure}

\begin{figure}[pthb]
\centering\includegraphics[width=8.cm,height=6cm]{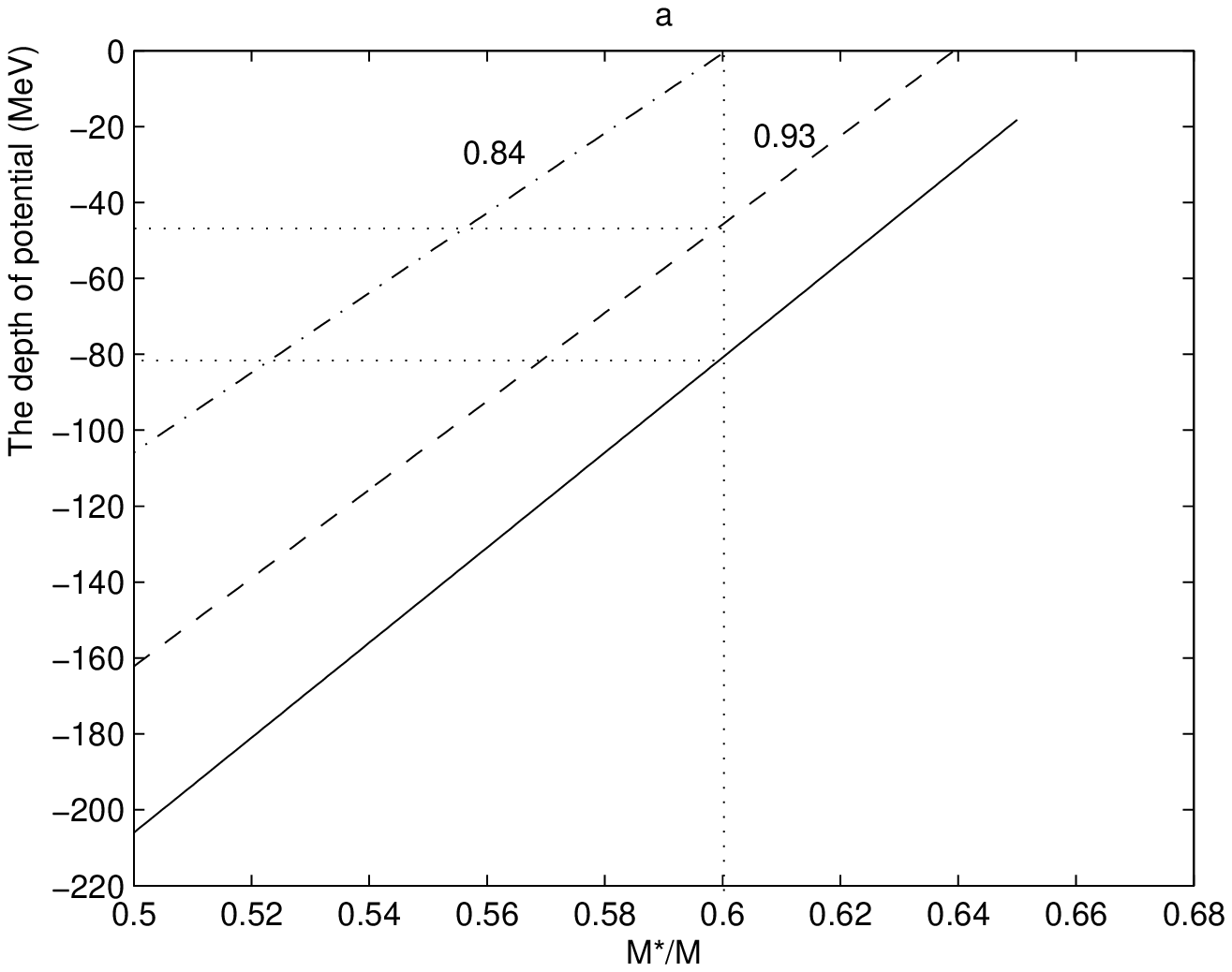}\includegraphics[width=8.cm,height=6cm]{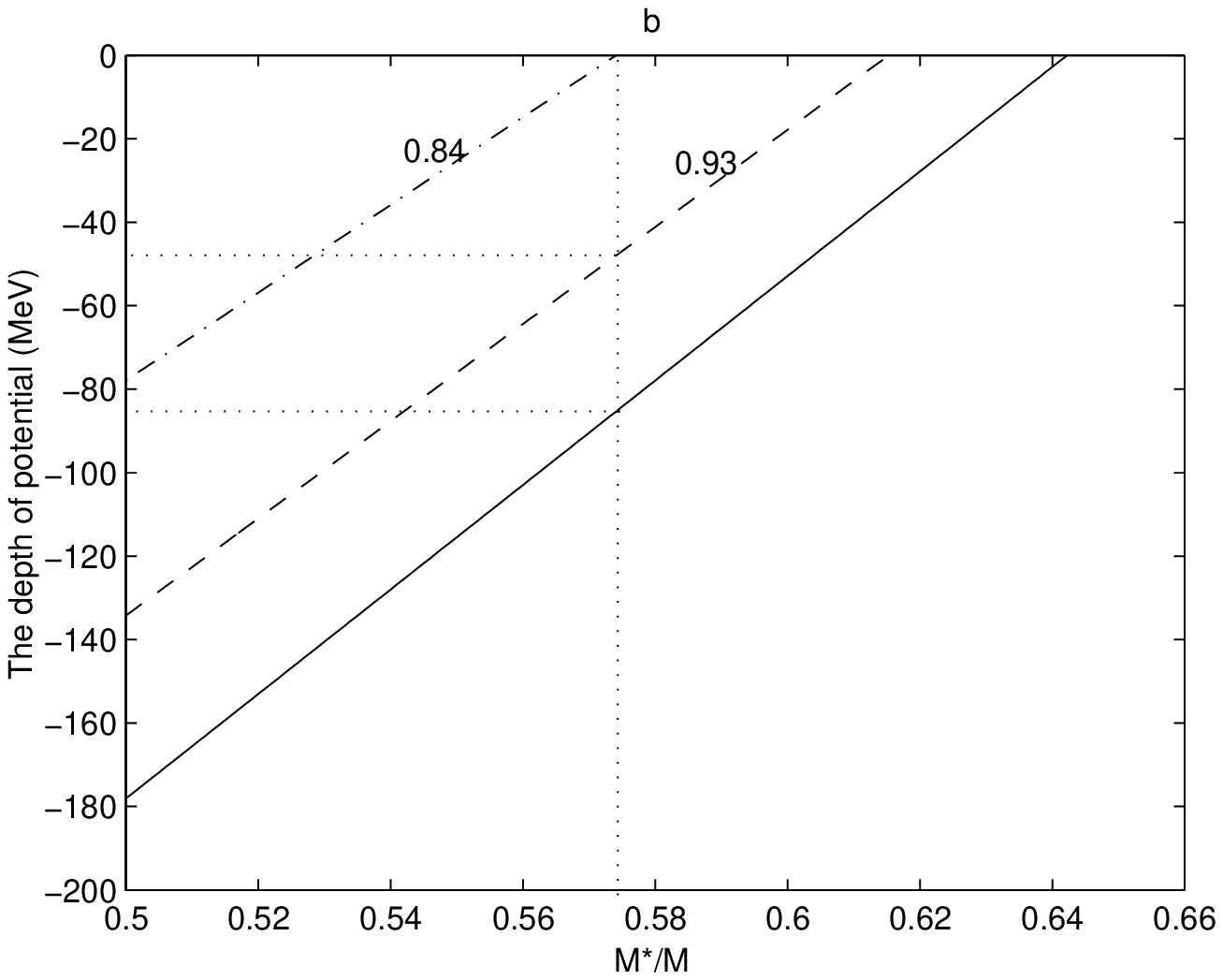}
\includegraphics[width=8.cm,height=6cm]{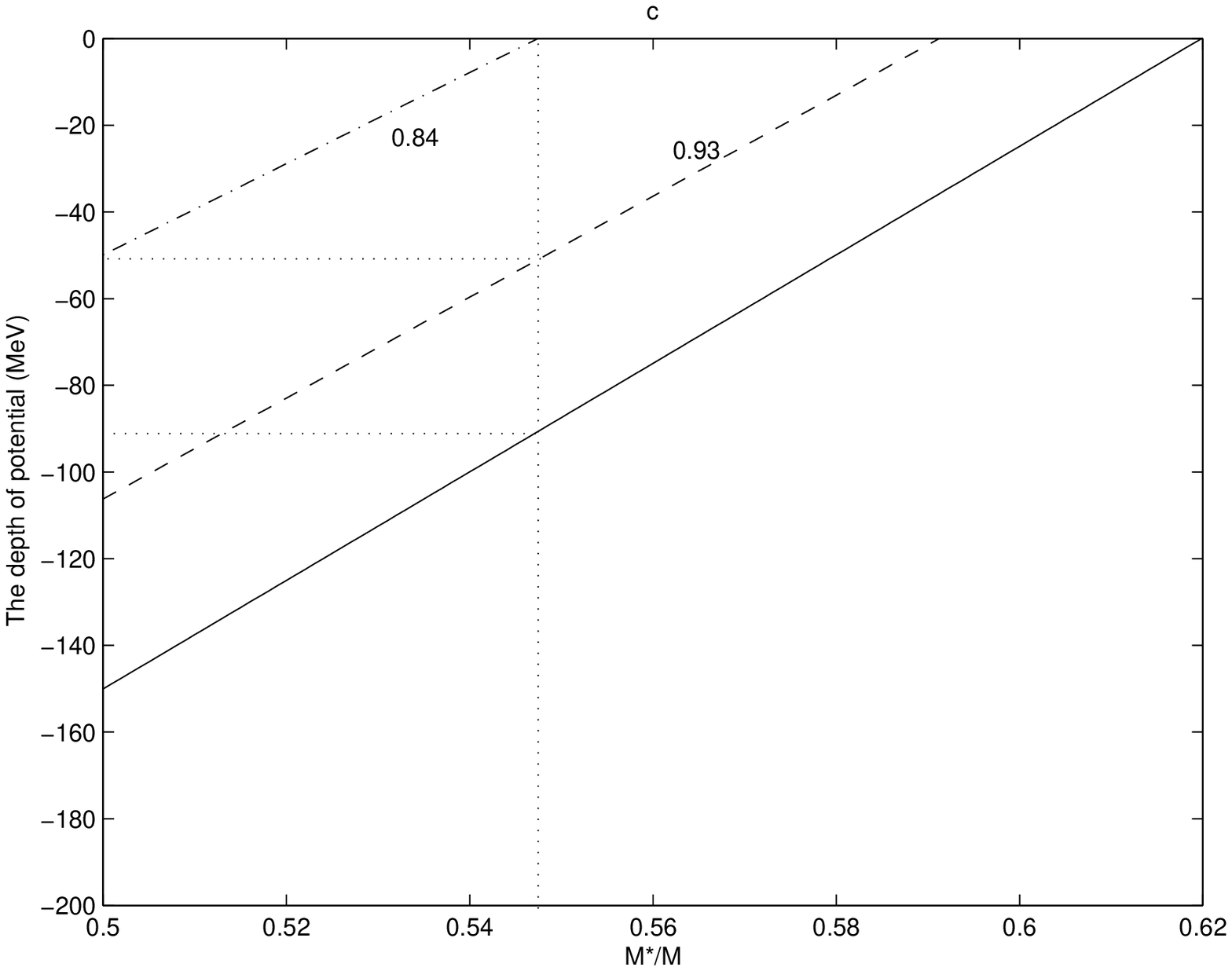}
 \caption{ The depth of potential for $\Theta^{+}$ in nuclear matter versus effective mass
 $M^{*}/M$ for fixed values of $U$ at the saturation
nuclear density. a,b and c is for $\varrho_{0}=0.15,0.16
,0.17$fm$^{-3}$ respectively. The dash dot line is for
$g_{\sigma}^{\Theta}=13.925\times0.84$, dash line is for
$g_{\sigma}^{\Theta}=13.925\times0.93$, and solid line is for
$g_{\sigma}^{\Theta}=13.925$. }\label{2}
\end{figure}

\end{document}